# National $PM_{2.5}$ and $NO_2$ Exposure Models for China Based on Land Use Regression, Satellite Measurements, and Universal Kriging

Hao Xu[1,2], Matthew J. Bechle[3], Meng Wang[4,5], Adam A. Szpiro[6], Sverre Vedal[5], Yuqi Bai[*1,2], Julian D. Marshall[*3]

1. The Ministry of Education Key Laboratory for Earth System Modeling, Department of Earth System Science, Tsinghua University, Beijing 100084, China
2. Joint Center for Global Change Studies(JCGCS), Beijing 100875, China
3. Department of Civil & Environmental Engineering, University of Washington, Seattle, Washington 98195, United States
4. Department of Epidemiology and Environmental Health, School of Public Health and Health Professions, University at Buffalo, Buffalo, NY, United States
5. Department of Environmental and Occupational Health Sciences, University of Washington, Seattle, Washington 98195, United States
6. Department of Biostatistics, University of Washington, Seattle, Washington 98195, United States

* Corresponding authors
E-mail addresses: jdmarsh@uw.edu (J. Marshall), yuqibai@tsinghua.edu.cn (Y. Bai).

ABSTRACT: Outdoor air pollution is a major killer worldwide and the fourth largest contributor to the burden of disease in China. China is the most populous country in the world and also has the largest number of air pollution deaths per year, yet the spatial resolution of existing national air pollution estimates for China is generally relatively low. We address this knowledge gap by developing and evaluating national empirical models for China incorporating land-use regression (LUR), satellite measurements, and universal kriging (UK). We test the resulting models in several ways, including (1) comparing models developed using forward stepwise linear regression vs. partial least squares (PLS) regression modeling, (2) comparing models developed with and without satellite measurements, and with and without UK, and (3) 10-fold cross-validation (CV), leave-one-province-out CV (LOPO-CV), and leave-one-city-out CV (LOCO-CV). Satellite data and kriging are complementary in making predictions more accurate: kriging improved the models in well-sampled areas; satellite data substantially improved performance at locations far away from monitors. Stepwise forward selection performs similarly to PLS in 10-fold CV, but better than PLS in LOPO-CV. Our best models employ forward selection and UK, with 10-fold CV $R^2$ of 0.89 (for both 2014 and 2015) for $PM_{2.5}$ and of 0.73 (year-2014) and 0.78 (year-2015) for $NO_2$. Population-weighted concentrations during 2014-2015 decreased for $PM_{2.5}$ (58.7 μg/m$^3$ to 52.3 μg/m$^3$) and $NO_2$ (29.6 μg/m$^3$ to 26.8 μg/m$^3$). We produced the first high resolution national LUR models for annual-average concentrations in China. Models were applied on 1 km grid to support future research. In 2015, more than 80% of the Chinese population lived in areas that exceed the Chinese national $PM_{2.5}$ standard, 35 μg/m$^3$. Results here will be publicly available and may be useful for epidemiology, risk assessment, and environmental justice research.

## 1. Introduction

Long term exposure to air pollutants such as fine particulate matter ($PM_{2.5}$) and nitrogen dioxide ($NO_2$) has been associated with many adverse health effects, including respiratory and cardiovascular diseases, and increased mortality.[1, 2] Epidemiological research on the health effects of air pollution exposure increasingly relies on high spatial resolution air pollution predictions.[3, 4] Land-use regression (LUR) and other empirical modeling approaches are useful tools to improve the accuracy of air pollution exposure estimates and to explore within-urban variability of outdoor air pollutants. LUR employs ground observations and geographic covariates to build a regression model and to estimate concentrations at locations without monitoring data, typically at a city-wide scale.[5] Variables corresponding to emission sources (e.g. traffic, population density, nearby pollutant emissions) and dispersion conditions (e.g. elevation, vegetative indices, meteorology) are often included in an LUR model. More recently, LUR and other geostatistical approaches have been used to model fine scale air pollution concentrations over large areas.[6-10] Unlike city-wide models, national LUR models typically rely on routine monitoring data instead of purpose-designed monitoring. As a rule of thumb, typically ~40-100 monitors are necessary to build a robust LUR model in relatively small areas,[11, 12] whereas large scale models have typically used ~300-900 monitors.[6-9] Satellite data and geostatistical methods such as universal kriging (UK) have been found to improve model performance when combined with LUR in large spatial scale models.[6, 8]

China is experiencing severe and widespread air pollution, reflecting rapid economic development and urbanization in recent years.[13] Public health studies conducted on national or regional scales have been critically important for China in advancing environmental policies to improve air quality.[14, 15] Because of limitations of data access and lack of publicly-available nationwide monitoring data prior to 2012, LUR models were rarely reported in China. Most reported studies focused on small-scale models that relied on limited number of GIS variables.[16-19] In recent years, satellite-data-driven national models have been emerging in China, which typically estimate the daily relations between a pollutant (*e.g.,* $PM_{2.5}$, $NO_2$) and satellite-derived aerosol optical depth (AOD) [20-23] or satellite-derived $NO_2$ [24]. These satellite-based models typically have relatively coarse spatial resolution (10 to 50 km) which may miss intra-urban variations. Incorporating local indicators of air pollution in an LUR framework could provide higher resolution predictions. Additionally, missing data due to cloud cover and weather conditions may increase uncertainty of daily satellite-based predictions.

Here we develop high-quality national LUR models for China that employ open-source GIS-derived land use and meteorological variables. Satellite data are incorporated to provide additional information especially at locations where monitors are sparse. Contributions of this paper to the literature include (1) first use of categorized points of interest (POI) data (e.g., gas stations, Chinese restaurants) and boundary-layer-height-averaged wind speed (BLHA-WS) as potential predictors in a national LUR model; (2) robust evaluation of satellite data and UK when these are incorporated in an LUR model in China, accounting for performance near and far from monitoring locations; (3) comparing model performance with forward stepwise regression and partial-least-squared (PLS) variable reduction methods; and (4) by focusing on long-term average concentrations, providing the high spatial resolution prediction maps of $PM_{2.5}$ and $NO_2$ ($1 \times 1$ $km^2$) in China, with evaluation of national, regional, and within-city variations. The publicly available predictions given here will be useful in advancing environmental and health studies in China, including in epidemiology and environmental health.

## 2. Materials and Methods

### 2.1 Monitoring Data

Daily mean ground-level $PM_{2.5}$ and $NO_2$ concentrations for two years (January 1st, 2014 to December 31st, 2015) were obtained from the China Environmental Monitoring Center (CEMC). [25] Measurements and quality control follow regulations of Chinese Ambient Air Quality Standards (GB 3095—2012) and Ambient Air Quality Index (AQI) technology (HJ 633—2012). Stations missing more than 25% of daily mean measurements for each pollutant were excluded; annual averages were calculated for each remaining monitor.

2.2 Geographical predictors

We employ a combination of point, buffer, and proximity based geographic variables resulting in 292 unique covariates. Details on each covariate, including the various buffer lengths we employ, are provided in Table S1 of the SI.

Road network data were extracted from OpenStreetMap[26], including all roads, major roads, secondary roads and railways. We calculate total length of road (all, major, and secondary) and railways within 16 buffer lengths from 100m to 10km. We also calculated distance to nearest major roads, secondary roads and railways.

The percentage of land cover types for eight categories was computed within 11 sizes of moving windows (from 300m to 30km). Land cover type datasets were derived from the Finer Resolution Observation and Monitoring of Global Land Cover dataset (FROM-GLC) with 30m resolution[27], resampled from sinusoidal projection to Albers projection system using nearest neighbor assignment.

Five types of POIs (i.e., gas stations, heat suppliers, polluting factories, bus stops and Chinese restaurants) were extracted using Amap API based on categories and keywords (see SI).[28] Categorized POIs may indicate local land uses that are not well captured by other variables and have been used in city-scale LUR, but to our knowledge have not previously been employed in a national LUR model. For example, Chinese restaurants are restaurants with Chinese-style cooking, which are important source of air pollution in China. [29] To capture both local and regional transport of air pollutants in China, we calculated POI counts using 22 buffer lengths from 100m to 50km.

Previous research suggested that remotely sensed fire count data could improve $PM_{2.5}$ prediction accuracy, and will have good prediction power when the buffer zone reach 50km.[30] To capture fire emissions, we used number of fire spots within 10 buffer lengths from 5km to 100km using Moderate-resolution Imaging Spectroradiometer (MODIS) Global Monthly Fire Location Product (MCD14ML)[31].

Other potential predictor variables included elevation (China 1km Digital Elevation Model data based on SRTM (Shuttle Radar Topography Mission))[32], population density (calculated from Landscan 2015[33]), Normalized difference vegetation index (NDVI) & Enhanced vegetation index (EVI) (derived from MODIS MOD13A3 monthly NDVI dataset[34]), and coordinates (x and y coordinate in China Albers Equal Area Conic coordinate system).

2.3 Meteorological data

Boundary layer height, temperature (at 2m), dew point temperature (at 2m), surface pressure and wind speed (at 10 m) were extracted from the European Reanalysis (ERA) Interim reanalysis data monthly means of daily means product.[35] We derived precipitation data from a 0.25°×0.25° interpolated observational product based on 2419 monitoring stations in China. [36] Relative humidity (RH) and BLHA-WS were also calculated (for details, see SI). [37] All meteorological data were averaged to annual means and re-sampled to 1km grid cells using bilinear interpolation.

## 2.4 Satellite-based air pollution data

To reduce the influence of possible deficiency of monitoring sites and improve the modeling accuracy, we incorporated satellite-based air pollution data. Satellite measurements of air pollution are derived from observations; in general, they reflect ambient conditions and therefore contributions from all emission sources. Satellite-based estimates have been previously developed for ground-level $PM_{2.5}$[38, 39] and $NO_2$[6, 9, 10]. We employed a 0.1° resolution global annual Satellite-Derived $PM_{2.5}$ product (http://fizz.phys.dal.ca/~atmos/martin/?page_id=140), which combined AOD retrievals from the NASA MODIS, Multi-angle Imaging SpectroRadiometer (MISR), and Sea-Viewing Wide Field-of-View Sensor (SeaWIFS) instruments with GEOS-Chem output to estimate ground-level $PM_{2.5}$. [40] There are two versions of the 0.1° satellite $PM_{2.5}$ data set, one as described above and a product calibrated to ground-based measurements using a geographically-weighted regression (GWR) [cite]. We employ the non-GWR data set, to avoid including satellite data calibrated to measurements from monitoring sites used for constructing our models. Previous studies suggest tropospheric $NO_2$ column data are sufficient to track spatial patterns in ground-level $NO_2$.[7, 41] We directly employed monthly mean Ozone Monitoring Instrument (OMI) tropospheric $NO_2$ column data with 0.125°resolution from the Derivation of OMI tropospheric $NO_2$ (DOMINO) product (version 1.0.2, collection 3; available at http://www.temis.nl). We calculated annual averages for 2014 and 2015 and then converted all of the satellite data into 1 × 1 $km^2$ grid cells using bilinear interpolation.

## 2.5 Statistical model building

We developed geostatistical models for mainland China for both $PM_{2.5}$ and $NO_2$ for the years of 2014 and 2015. We used R X64 3.4.0 'stat' and 'gstat' packages.[42] To determine the added value of UK and satellite data, we developed models with and without UK, and with and without satellite data.

### 2.5.1 Stepwise regression

At the first stage, we used conventional multivariable linear regression based on variable selection approach. For each pollutant ($PM_{2.5}$, $NO_2$) and year, we explored four potential regression models: satellite $PM_{2.5}$ included; satellite $NO_2$ included; both satellite $PM_{2.5}$ and $NO_2$ included; and, without satellite data. We followed conventional supervised forward stepwise linear regression, with 290 to 292 independent variables as inputs in our forward selection.[43] (Separately, we also tried building models using PLS; see below for details.) Briefly, the independent variable most correlated with the dependent variable was added to the model. In subsequent steps, the remaining variable that gave the largest improvement in adjusted $R^2$ was added to the model if (1) the variance inflation factor (VIF, a check for multi-collinearity) of the variable was less than 5; (2) the *p*-value of the variable was less than 0.05; and (3) the direction of existing variables in the model did not change. This procedure was repeated until the increase in adjusted-$R^2$ for an additional variable was less than 0.5% or no variable met the aforementioned criteria.

We allowed multiple buffer lengths per variable (e.g. major road length) to be selected into the model as long as they follow our criteria.[44]

We also tried alternative metrics (e.g., F value, 10-fold $R^2$, adjusted $R^2$, Akaike Information Criterion (AIC)), and alternative algorithms including forward selection and backward selection, to see whether those attributes strongly influenced results from the model-building process.

Monte Carlo iterations were used to evaluate the over-fitting risk; we simulated reduced data availability via Monte Carlo sampling (500 iterations per number-of-monitors) at lower number of monitors, and then compared model fitted and cross-validated model performance. Other regression diagnostic tests included checking the normality of residuals, heteroscedasticity, and spatial autocorrelation of residuals using Moran's I.

### 2.5.2 Universal kriging

At the second stage, we incorporated a spatial smoothing approach (UK). Kriging will account for spatial autocorrelation in the model residuals. We leveraged a first-order polynomial function in UK as the external drift, whose independent variables were obtained from the forward selection stage. We modeled spatial dependence using the exponential variogram model. The formula for the models with UK can be given by equation (1):

$$Y(s) = (\beta_0+\beta_0') + (\beta_i+\beta_i')\times X(s) + \varepsilon(s) \quad (1)$$

where Y(s) denotes the annual mean concentrations of the pollutant at the monitoring locations, $\beta_0$ and $\beta i$ are the intercept and coefficients of the linear function in the first stage, $\beta_0'$ and $\beta_i'$ are the adjusting intercept and coefficients from the external drift in the UK function, X(s) denotes the matrix of spatially varying independent variables selected from the first stage, ε(s) indicates spatially varying residuals modeled by UK.

### 2.5.3 Partial Least Squares (PLS) regression

Some previous empirical models for the US were built using partial least squares (PLS) combined with UK.[6, 8] PLS reduces the dimensions of the many predictor variables and avoids having to use a variable selection procedure. To evaluate the comparative performance of variable-selection-based models and PLS-based models, we also built PLS models with and without satellite data, and with and without UK, using all potential predictors for 2015. Selection of PLS components were based on 10-fold CV, using the R command selectNcomp (result: the most parsimonious model not significantly worse than the model with global minimum root mean square error of prediction is selected, see Figure S6 in SI). Details of our model building approaches are described elsewhere.[6] Briefly, the satellite data were used directly as a covariate in the PLS procedure and spatially varying PLS components were used in building UK models.

## 2.6 Cross-Validation and Model Assessment

Our core model evaluation mainly used two types of cross-validation approaches: conventional 10-fold cross-validation (10-fold CV, as default CV method), and Leave-One-Province-Out cross-validation (LOPO-CV). For conventional cross-validation, all monitoring sites were randomly divided into 10 groups. Nine groups were then used to train the model and the remaining one group to test the model. This process was repeated 10 times, until all the groups were tested, resulting in "out-of-sample" predictions at all monitoring sites.

Most monitors are in cities, and often are somewhat near to (in the same city as) other monitors. To explore model performance at locations without a nearby monitor, we performed LOPO-CV on each model, wherein we consecutively exclude monitors from one province.

To further address the city-scale performance of our national models, we additionally conducted Leave-One-City-Out cross-validation (LOCO-CV), wherein we exclude all monitors from a city during model-building, and then compare model results against (held-out) monitoring data for that city.

Statistics from 10-fold CV and LOPO-CV/LOCO-CV used to assess model performance include mean-square-error-based $R^2$ (assessing deviation around the 1:1 line)[45] and root-mean-squared-error (RMSE). In addition, in order to demonstrate model differences spatially, we calculated and mapped differences between the national predictions of the different models.

## 3. Results

### 3.1 Model results and comparison

For 2014 and 2015, respectively, the number of monitors that met our inclusion criteria was 893($NO_2$)/902($PM_{2.5}$) and 1418($NO_2$)/1419($PM_{2.5}$). Descriptive statistics for those input data are in Table S8 in SI. The large difference in the number of monitors between 2014 and 2015 is the result of a rapid monitoring network expansion in China since 2012; there were a total of 944 monitors in 2014 and 1494 in 2015. Figure 1 presents the $R^2$ of the 2015 models based on variable selection method. Table 1 shows the summary of all the models built for 2015 (2014 model result are presented in Table S9 in the SI).

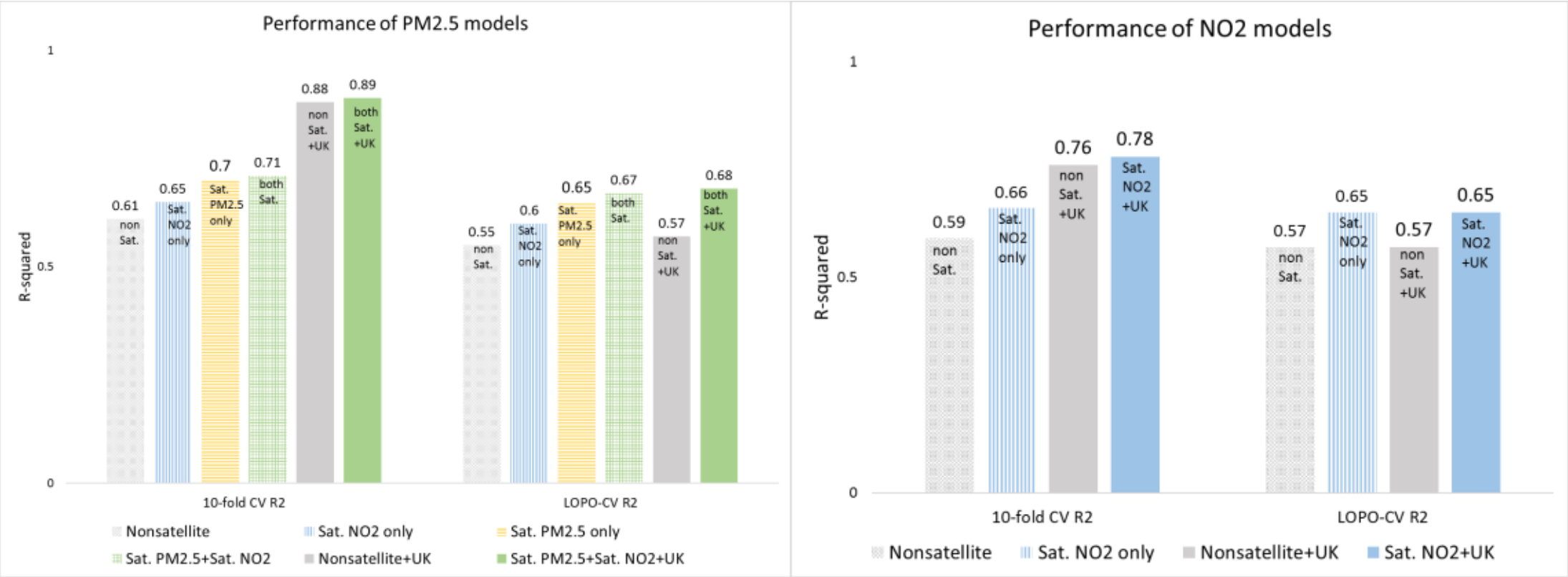

**Figure 1.** Model performance for year-2015 models. (Sat.=satellite) Other statistics (e.g., RMSE; results for year-2014) are in SI.

**Table 1** Summary of all the models built for 2015, (VS= Variable Selection, PLS= Partial Least Square, Sat.=satellite, UK =Universal Kriging, LOPO=Leave One Province Out).

| Model | VS | PLS | Sat. PM2.5 | Sat. NO2 | UK | 10-fold CV $R^2$ | LOPO CV $R^2$ | 10-fold RMSE (μg/m³) | LOPO RMSE (μg/m³) |
|---|---|---|---|---|---|---|---|---|---|
| $PM_{2.5}$-1 | ✓ | | | | | 0.61 | 0.55 | 11.4 | 12.3 |
| $PM_{2.5}$-2 | ✓ | | ✓ | | | 0.70 | 0.65 | 10.2 | 10.9 |
| $PM_{2.5}$-3 | ✓ | | | ✓ | | 0.65 | 0.60 | 10.9 | 11.7 |
| $PM_{2.5}$-4 | ✓ | | ✓ | ✓ | | 0.71 | 0.67 | 9.9 | 10.6 |
| $PM_{2.5}$-5 | ✓ | | | | ✓ | 0.88 | 0.57 | 6.3 | 12.1 |
| $PM_{2.5}$-6 | ✓ | | ✓ | ✓ | ✓ | 0.89 | 0.68 | 6.3 | 10.4 |
| $PM_{2.5}$-7 | | ✓ | ✓ | ✓ | | 0.70 | 0.64 | 10.0 | 11.1 |
| $PM_{2.5}$-8 | | ✓ | ✓ | ✓ | ✓ | 0.89 | 0.66 | 6.1 | 10.8 |
| $NO_2$-1 | ✓ | | | | | 0.59 | 0.57 | 7.9 | 8.1 |
| $NO_2$-2 | ✓ | | | ✓ | | 0.66 | 0.65 | 7.2 | 7.4 |
| $NO_2$-3 | ✓ | | | | ✓ | 0.76 | 0.57 | 6.1 | 8.1 |
| $NO_2$-4 | ✓ | | | ✓ | ✓ | 0.78 | 0.65 | 5.9 | 7.4 |
| $NO_2$-5 | | ✓ | | ✓ | | 0.63 | 0.57 | 7.6 | 8.1 |
| $NO_2$-6 | | ✓ | | ✓ | ✓ | 0.76 | 0.60 | 6.1 | 7.9 |

### 3.1.1 Variable-selection models

Detailed descriptions of all the variable-selection models are in Table S6 and Table S7 in SI. $PM_{2.5}$ models explained 69%-76% variation in 2014 and 62%-71% in 2015. Including satellite-derived $PM_{2.5}$ and $NO_2$ together greatly improved 10-fold CV $R^2$ for $PM_{2.5}$ (e.g., 16% improvement in 2015) compared to non-satellite model. The prediction ability of satellite-derived $PM_{2.5}$ is slightly better than satellite-derived $NO_2$ when only using one set of satellite data. The best variable-selection model ($PM_{2.5}$-4) suggest that agricultural emission source (percentage of cropland), indirect traffic/urbanization indicators (number of gas stations/bus stops, road length) and meteorological conditions (BLHA-WS, RH) are important predicting factors to $PM_{2.5}$ models.

Similarly, including $NO_2$ satellite data substantially improved the $NO_2$ model CV $R^2$, e.g., by 12 percentage points in 2015, whereas satellite-derived $PM_{2.5}$ was not selected into any $NO_2$ model (hence, it is not displayed for $NO_2$ models in Figure 1). Model CV $R^2$'s were lower for $NO_2$ than for $PM_{2.5}$ (0.61 [$NO_2$] vs. 0.76 [$PM_{2.5}$] in 2014; 0.66 [$NO_2$] vs. 0.71 [$PM_{2.5}$] in 2015). Key predictor variables for the $NO_2$ models included urbanicity (percentage of impervious surfaces, percentage of forest, number of heating suppliers), indirect traffic/urbanization indicators (number of gas stations/bus stops) and meteorological conditions (BLHA-WS). Model buffer lengths were generally smaller for $NO_2$ than $PM_{2.5}$, consistent with $PM_{2.5}$ being a more regional pollutant than $NO_2$.

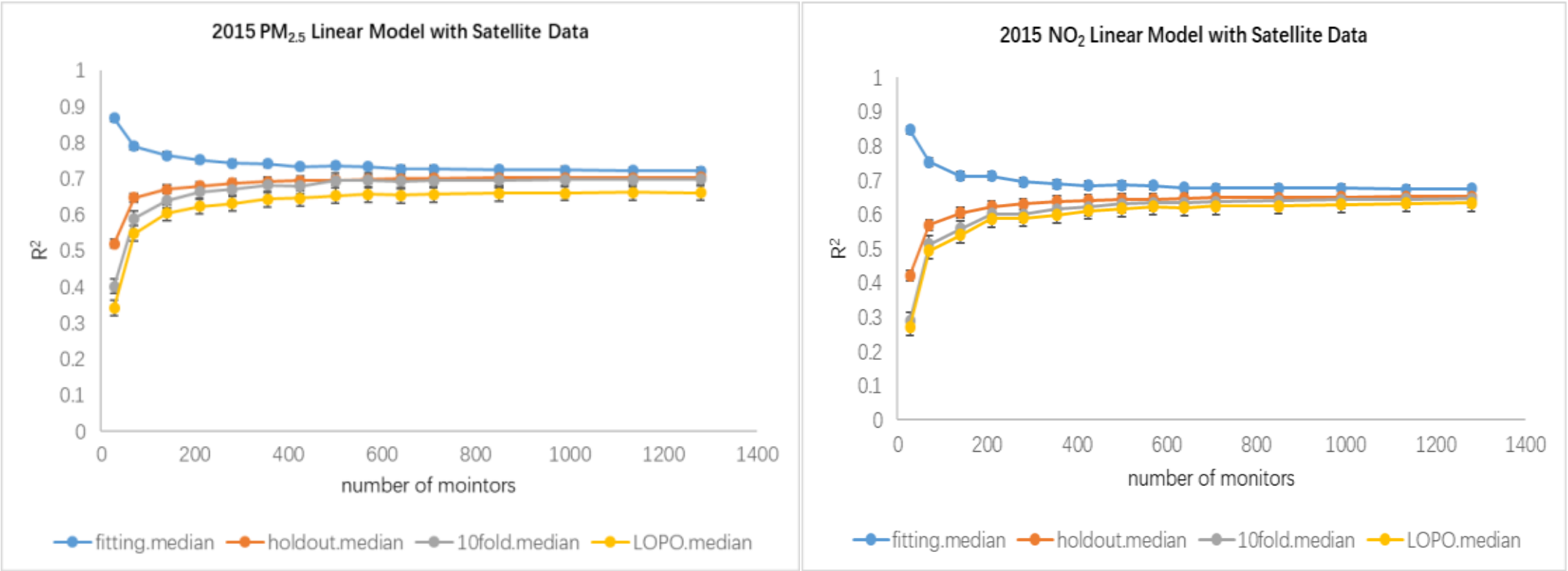

**Figure 2.** Median and interquartile range $R^2$ for Monte Carlo random sampling for n training monitors employed in model building (left: 2015 $PM_{2.5}$ LUR model with satellite data; right: 2015 $NO_2$ LUR model with satellite data). Fitting $R^2$ used n monitors to fit the model; holdout $R^2$ used n monitors only to build the model, the rest of monitors are used to test the model, and the $R^2$ calculated using test sets; 10-fold $R^2$ used n monitors to build and test model using 10-fold CV; LOPO $R^2$ used n monitors to build and test model using LOPO-CV.

Results from our test to evaluate the potential for over-fitting in our year-2015 models is presented in Figure 2 (year-2014 results are in Figure S2 in SI). The evaluation metric (model $R^2$) converges at ~ 400-450 monitoring sites, suggesting that the number of monitoring sites in our models were more than sufficient, with little risk of over-fitting. The difference between 10-fold CV $R^2$ and LOPO-CV $R^2$ is slightly larger in $PM_{2.5}$ models than in $NO_2$ models (e.g. 0.71 vs. 0.67 for $PM_{2.5}$ and 0.66 vs 0.65 for $NO_2$ in 2015), suggesting that $NO_2$ predictors are slightly more capable of capturing spatial variance at locations far away from training samples, but differences are modest. We also classified model $R^2$ by rural, suburban and urban areas based upon population density using our best performing linear models ($PM_{2.5}$-4 and $NO_2$-2) (see Figure S3 in SI). The $PM_{2.5}$ linear models yielded the best predictions in urban areas (CV $R^2$: 0.79 in 2014, 0.73 in 2015); $NO_2$ linear models gave best predictions in rural areas (CV $R^2$: 0.64 in 2014, 0.71 in 2015).

### 3.1.2 Kriging models

Figure 1 also presents the results of model performance with addition of UK. Incorporating UK improved the 10-fold CV $R^2$ for $PM_{2.5}$ and $NO_2$; increases were 0.17 to 0.27 for non-satellite models, 0.12 to 0.18 for satellite models. The differences in 10-fold CV $R^2$ between kriging models with and without satellite data were small (ranging from 0.00 to 0.03), however, under LOPO-CV, kriging models were improved with the addition of satellite data ($R^2$ increased 0.08 to 0.11). Under 10-fold CV, the best models ($PM_{2.5}$-6 and $NO_2$-4) consistently included satellite data with UK.

### 3.1.3 Comparison with PLS models

Using 10-fold CV $R^2$, model performance for $PM_{2.5}$ and $NO_2$ was similar for PLS models ($R^2$: 0.89 [$PM_{2.5}$], 0.76 [$NO_2$] in 2015) as for conventional variable-selection with UK. However, using LOPO-CV, PLS models with satellite data and UK had $R^2$ values of 0.66 ($PM_{2.5}$) and 0.60 ($NO_2$) – slightly worse than similar conventional models (0.68 and 0.65 for $PM_{2.5}$ and $NO_2$, respectively). Our PLS models used from 6 to 8 components. Because PLS models performed no better than the conventional variable-selection models (see table 1 and table S9 in the SI), and in the case of LOPO-CV were worse, we chose the variable-selection model with satellite data and UK as our core model.

3.2 Model predictions and assessments

Figure 3 shows our annual prediction maps for China based on our best performed models ($PM_{2.5}$-6 and $NO_2$-4) in 2015 (for 2014, see Figure S11), consisting of 9.6 million 1×1 $km^2$ grid cells. The most polluted areas for $PM_{2.5}$ were in the Beijing-Tianjin-Hebei urban agglomeration: predicted annual-average concentrations were above 85 $\mu g/m^3$ in 2014 and above 75 $\mu g/m^3$ in 2015. In eastern and northern China, $PM_{2.5}$ concentrations were similar (above 60 $\mu g/m^3$ in 2014 in most areas). Regions in central and western China including Hunan and Hubei provinces and the Sichuan basin also exhibit comparatively higher concentrations of $PM_{2.5}$. Also, $PM_{2.5}$ concentrations were high in the southern part of Xinjiang autonomous region where transported dust from deserts may be a major source. For $NO_2$, the most polluted areas are urban areas, especially the Beijing-Tianjin-Hebei urban agglomeration, Shandong province, the Yangtze River Delta and the Pearl River Delta. These regions are more economically developed and densely populated and have more industrial sources.

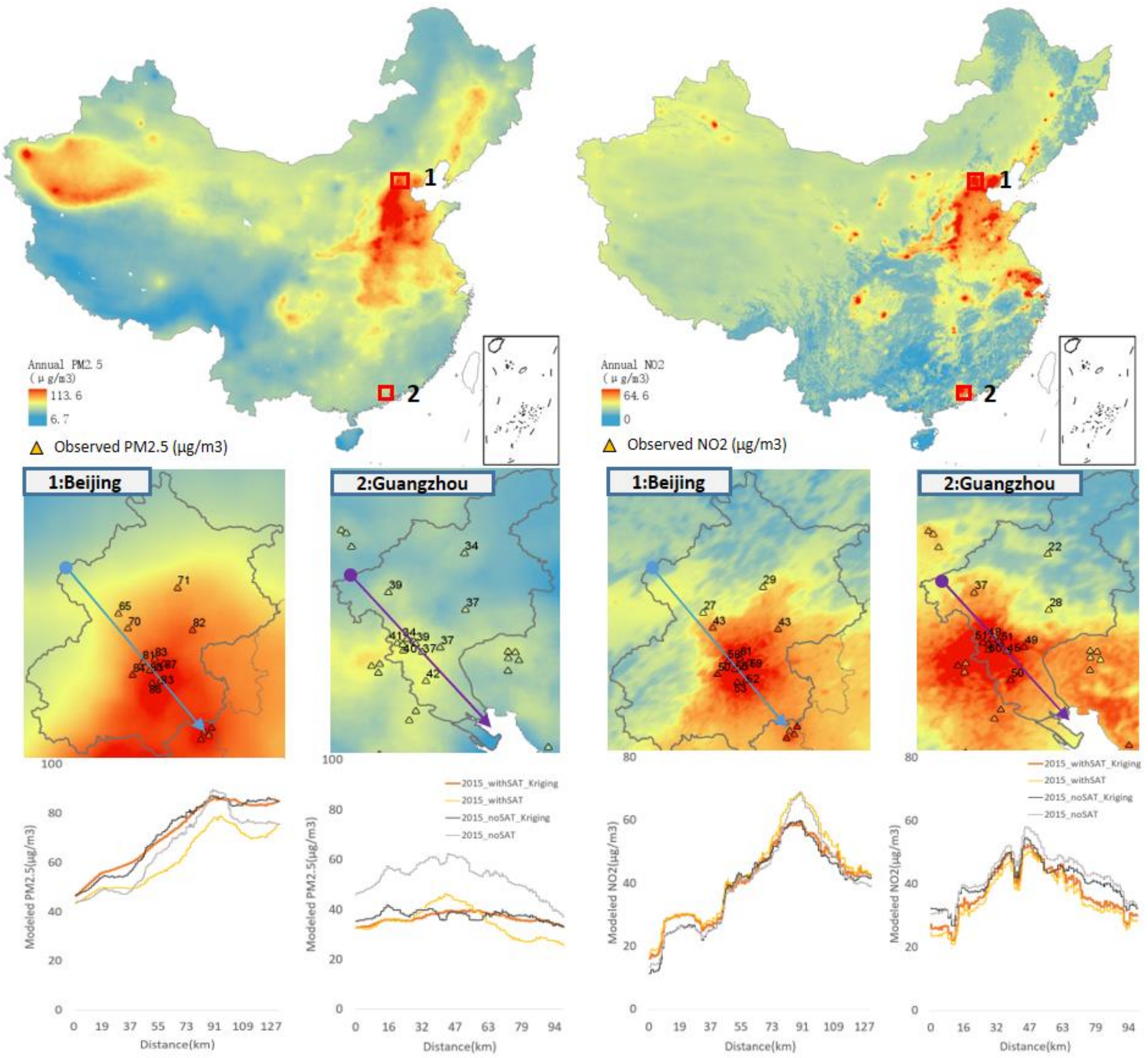

**Figure 3.** National (top) and city-level (middle) predictions derived from our best models ($PM_{2.5}$-6 and $NO_2$-4) in 2015 and profile plots of concentrations in two major cities in China based on four different models. (left:$PM_{2.5}$ right:$NO_2$). Profile plots are derived from 1×1 $km^2$ estimates along the transect shown for each city. Monitor locations are indicated with triangle symbols in city-level maps along with corresponding monitor concentration.

Maps for predictions from the various approaches (see Figure S8 in the SI) suggest consistent patterns in spatial variation of the pollutants. However, the inclusion of satellite data has a more regional impact, and typically provides information in areas with few monitors (e.g. Xinjiang, Tibet and Northeastern China). Kriging mainly creates adjustments in urban areas, where monitor density is greater.

Figure 3 also shows year-2015 spatial predictions from the four models (with and without satellite data; with and without UK) along transects across two major cities in Northern and Southern China (Beijing and Guangzhou). For $PM_{2.5}$ models, although the addition of satellite data and kriging resulted in better model performance, the spatial concentration gradients became smoother with some potential loss of spatial variation. For predicted concentrations, differences between satellite and non-satellite models is comparatively smaller for $NO_2$ than for $PM_{2.5}$. Within-city variations are greater for $NO_2$ predictions than $PM_{2.5}$ predictions. The patterns described here are for 2015; patterns for 2014 were similar (see Figure S11 in SI).

We used predictions derived from CV results of our best national models ($PM_{2.5}$-6 and $NO_2$-4) to calculate citywide $R^2$ and RMSE for 10 major cities. We selected the 10 cities with the largest number of monitors. 10-fold CV and LOCO-CV were used for model evaluation. As shown in Table 2, in 2015, 10-fold CV $R^2$ of $PM_{2.5}$ models ranged from 0.01 to 0.79, with RMSE ranging from 3.9 to 6.8. Although city-scale $R^2$ of $PM_{2.5}$ model were relatively poor (most cities are below 0.10), the RMSE values in these cities were excellent (most below 5.0); that result suggests that the $PM_{2.5}$ model is accurately predicting (average) concentrations in each city, but that within-city spatial variability is either too low or not well captured by the model. In contrast, the $NO_2$ model in most cities performed reasonably well (eight cities had a 10-fold CV $R^2 > 0.50$). LOCO-CV reflected model performance when models were built excluding the monitors in the specific city; for cities with monitors, performance for our final models ($PM_{2.5}$-6 and $NO_2$-4) will generally be better than LOCO-CV results. Under LOCO-CV, the RMSE values were higher in Northern cities such as Beijing,

Tianjin and Shenyang, where concentrations of air pollutants were also relatively high.

**Table 2** City-scale performance of final national models (2015)

| *City* | *N* | *$PM_{2.5}$* | | | | *$NO_2$* | | | |
|---|---|---|---|---|---|---|---|---|---|
| | | 10-fold $R^2$ | 10-fold RMSE (μg/m$^3$) | LOCO $R^2$ | LOCO RMSE (μg/m$^3$) | 10-fold $R^2$ | 10-fold RMSE (μg/m$^3$) | LOCO $R^2$ | LOCO RMSE (μg/m$^3$) |
| *Chongqing* | 17 | 0.06 | 6.8 | 0.24 | 7.1 | 0.65 | 6.4 | 0.76 | 7.8 |
| *Beijing* | 12 | 0.79 | 4.1 | 0.78 | 10.0 | 0.77 | 6.0 | 0.83 | 8.6 |
| *Tianjin* | 11 | 0.18 | 6.0 | 0.21 | 15.0 | 0.19 | 4.9 | 0.07 | 14.3 |
| *Hangzhou* | 11 | 0.76 | 5.4 | 0.78 | 5.9 | 0.78 | 6.3 | 0.83 | 6.8 |
| *Shenyang* | 11 | 0.55 | 4.3 | 0.44 | 8.6 | 0.43 | 4.5 | 0.41 | 7.9 |
| *Guangzhou* | 10 | 0.04 | 3.9 | 0.10 | 4.0 | 0.79 | 5.2 | 0.83 | 5.5 |
| *Wuhan* | 10 | 0.09 | 4.4 | 0.21 | 6.6 | 0.77 | 5.4 | 0.85 | 7.5 |
| *Changchun* | 10 | 0.34 | 4.2 | 0.54 | 4.9 | 0.71 | 6.7 | 0.79 | 7.5 |
| *Changsha* | 10 | 0.04 | 4.7 | 0.17 | 5.0 | 0.57 | 3.9 | 0.71 | 4.0 |
| *Shanghai* | 9 | 0.02 | 4.8 | 0.30 | 8.6 | 0.43 | 4.5 | 0.41 | 7.9 |

Figure 4 shows cumulative exposure nationally for $PM_{2.5}$ and $NO_2$, based on best performing models ($PM_{2.5}$-6 and $NO_2$-4). For $PM_{2.5}$, more than 90% of people in year-2014 lived in locations that exceed China's national standard, 35 μg/m$^3$ (same as WHO IT1); this number reduced to 83% in 2015. Average concentrations were above 75 μg/m$^3$ $PM_{2.5}$ for >20% of people, and above 40 μg/m$^3$ $NO_2$ for >20% of people.

**Figure 4.** Cumulative exposure for $PM_{2.5}$ and $NO_2$ based on the best performing models ($PM_{2.5}$-6 and $NO_2$-4). For context, air quality guidelines (AQG) and interim targets (IT1-3) from the World Health Organization are shown.

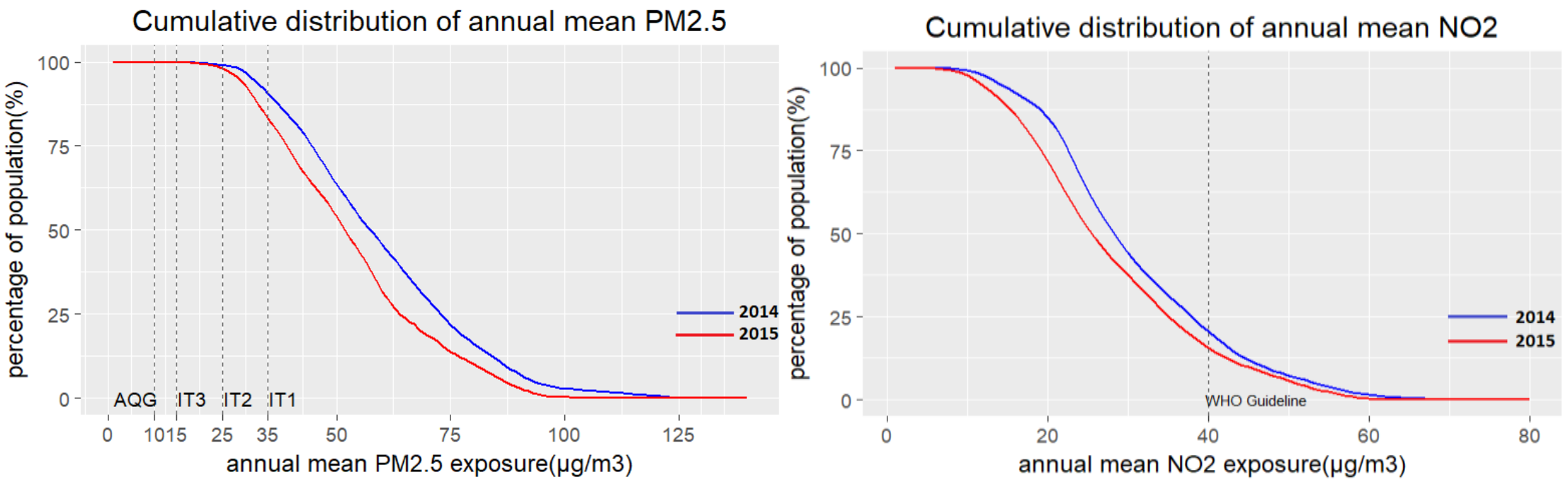

## 4. Discussion

Our research developed and rigorously tested national prediction models for $PM_{2.5}$ and $NO_2$ in China using large open-source datasets and state-of-the-art modeling. Factors influencing air quality may vary by year; we built separate models for each year (2014; 2015) and pollutant. Our final models ($PM_{2.5}$-6 and $NO_2$-4) incorporated satellite data and UK and exhibited good predictive power (10-fold CV $R^2$: 0.89 [$PM_{2.5}$], 0.73 to 0.78 [$NO_2$]).

4.1 Model performance

We compared model performance of our variable-selection-based models with PLS-based models. Although PLS obviates the need for variable selection and deals with multicollinearity, it is computationally intensive for making national predictions since all of the geographic variables need to be used for extracting the individual PLS components. In addition, not all potential variables are correlated with the dependent variables ($PM_{2.5}$ and $NO_2$); since PLS uses all of the variables, this aspect raises potential concern of overfitting. Further, it is not straightforward to demonstrate the contribution of each geographic variable to overall model predictions. We found that PLS (alone or combined with UK) performs similarly to, or in some cases not as well as, multivariate linear regression combined with UK.

Performance of our final $PM_{2.5}$ models ($PM_{2.5}$-6) were comparable to those reported from previous studies in the US (10-fold CV $R^2$ 0.89 vs. 0.88), [8] while performance of the $NO_2$ models was slightly worse than that reported from the US (10-fold CV $R^2$ 0.78 vs. 0.85); [6] potential explanations include that in China relative to the US monitoring sites might be located more unevenly, or explanatory variables we employed are less relevant to pollution, or pollution may be generally less correlated with land use.

Most existing large scale empirical models in China are daily models based on constructing relations between satellite data and ground observations (details regarding existing models are in table S11 in SI). Reported overall $R^2$'s of these models are 0.62 to 0.80. Very few publications reported annual $R^2$. Zhan et al. used machine learning to model daily $PM_{2.5}$ in 2014 at 50km grid cells in China, yielding an annual $R^2$ of 0.84 based on 10-fold CV. [46] Xue et al. estimated daily $PM_{2.5}$ in 2014 with 0.1° spatial resolution combined satellite data and CMAQ model output, yielding an annual $R^2$ of 0.87 based on site-based CV.[38] Zhan et al. predicted 0.1° spatial resolution daily $NO_2$ from 2013 to 2016 using random forest and spatiotemporal kriging, yielding an annual $R^2$ of 0.68.[24] Our parsimonious models with relatively low computational cost had comparable performance without requiring complex algorithms. Furthermore, our method provided very fine scale predictions at 1km resolution while explicitly assessing the relationships between the pollutants and land use variables. Prior research has applied the LUR approach for smaller regions in China; for example, Yang et al. developed a regional LUR for the Pearl River Delta region.[47]

4.2 Contribution of satellite data and UK

When including all potential predictors, both satellite $PM_{2.5}$ and satellite $NO_2$ were consistently selected into $PM_{2.5}$ models, while only satellite $NO_2$ was consistently selected into $NO_2$ models. Satellite data substantially improved LOPO-CV performance of the models, suggesting that satellite data provides additional spatial information on air pollutant concentrations that land use and meteorological variables could not provide. It is worth noting that satellite $NO_2$ could compliment satellite $PM_{2.5}$ data in $PM_{2.5}$ models, serving as a proxy for polluted urban plumes, however, satellite $PM_{2.5}$ has low correlation with $NO_2$ concentration; satellite $PM_{2.5}$ was never selected into our $NO_2$ models. According to model $R^2$, satellite data contribute more to improving models for $PM_{2.5}$ than for $NO_2$, a finding consistent with results reported previously for Europe.[48] Previous national models in the US show an increase in $R^2$ of 0.22 for a $PM_{2.5}$ model [49] and an increase in $R^2$ of 0.12 for an $NO_2$ model [6] by incorporating satellite data.

As shown in table 1 and table S9, UK made substantial improvement in 10-fold CV performance, with little distinction in performance between UK models with and without satellite data. Under extreme conditions like LOPO-CV, performance of all models was reduced owing to prediction errors in large unmonitored areas, however, UK models with satellite data performed better in LOPO-CV than UK models without satellite data. This suggests that models with UK may mask the importance of satellite data (or possibly other regional predictors) when evaluated with 10-fold CV, and highlights the importance of alternative CV evaluation such as LOPO-CV. These findings are consistent with a previous study in the US,[6] though the improvement from UK (0.12-0.18 increase in 10-fold CV $R^2$ for satellite models) was greater than for the US (0.04 increase in 10-fold CV $R^2$).

The overall performance of the $PM_{2.5}$ models was consistently better than the $NO_2$ models, perhaps because our predictors were better at explaining regional concentrations of a more regional pollutant such as $PM_{2.5}$ rather than a pollutant with more local sources, such as $NO_2$. Although all of our models yielded reasonable $R^2$ and RMSE, there was systematic underestimation for both $PM_{2.5}$ and $NO_2$, especially when measured concentrations were high (see scatter plots in Figure S3 and S4 in SI). Possible reasons for this could be some inadequacy of our predictors in national scale modeling and possible non-linear relationships between the dependent variables and the predictors. This finding is unsurprising; most models are better at detecting central tendencies than at accurately predicting extremes.

### 4.3 Variable selection

Because variables selected into our final models tended to be relatively stable across the different approaches (described in Section 2.5.1), we chose a relatively simpler and more conventional approach to select variables (forward selection based on adjusted-$R^2$).

Many previous LUR studies used emission data as an important predictor.[7-9] Since we lacked emission data and detailed information on pollution sources, we employed some alternative variables from open-source datasets. We used number of fire spots to reflect pollution from biomass burning, and number of different types of POIs to reflect industrial source pollution, heating suppliers, urban transportation and cooking fumes. POI data was an important predictor in our final models (e.g., gas stations, bus stops, heating suppliers), suggesting that POI data from online mapping services (e.g., Amap, Google Maps) may provide useful local-scale information for national-scale LUR. Prior studies have reported that meteorological factors like wind speed[50], wind direction[51], precipitation[7] and boundary layer height[52] are useful for predicting $PM_{2.5}$ and $NO_2$ concentrations. Here, we incorporated BLHA-WS to represent diffusion conditions of air pollutants and found it contributed in both $PM_{2.5}$ and $NO_2$ models. In our non-satellite LUR models, percentage of cropland and number of gas stations consistently entered in $PM_{2.5}$ models, while number of gas stations and percentage of impervious land entered in $NO_2$ models. Crop land could be a non-negligible source of $PM_{2.5}$ when generated from ammonia, acid gases and straw burning and (for primary $PM_{2.5}$) dust.[53, 54] Variables with larger buffer lengths were more likely to be selected into $PM_{2.5}$ models, while buffer lengths of variables selected into $NO_2$ models were comparatively smaller. This may be because $PM_{2.5}$ is a more regional pollutant affected by long-range transport: variables with large buffer sizes could reflect more regional transport, while $NO_2$ concentrations are more likely to be affected by local pollution sources.

4.4 Cross validation

Distance between a test-set monitor and its nearest training-set neighbor for LOPO-CV ranged from 28 km to 1,454 km with a mean (median) value in 2015 of 184 km (148.7 km); for 10-fold CV the same value (distance between test monitor and nearest training-set neighbor) ranged from 0.3 km to 350 km with a mean (median) of 8.9 km (3.8 km) (see also table S4). Figure S13 indicates that most people live in areas less than 50km from the nearest monitor. That result indicates that LOPO-CV is a more extreme (more stringent) test of model performance than would be applicable to most people in mainland China. Average model performance across the population should fall between 10-fold CV performance and LOPO-CV performance. Our findings indicate that 10-fold CV may overestimate model performance at locations far from monitors, whereas LOPO-CV likely underestimates model performance for most people. Table 1 and Table S9 show that 10-fold CV performance was consistently better than LOPO-CV performance. This finding also suggests that performance of both kriging and non-kriging models is reduced in unmonitored regions. Previous studies have also implemented some approaches that non-randomly select cross-validation groups, such as spatially clustered cross-validation[6] and isolated-site cross-validation[38], which tend to select the test-set to be far away from the training-set. Based on the characteristic of the monitoring sites distribution in China, we chose to use provinces or cities as fixed groups to conduct cross-validation, which is convenient for quantitatively analyzing differences in model performance of the CV methods and in evaluating model performance at the province or city scale. [55]

### 4.5 Within-city variation

We were able to assess within-city variation in concentrations using our national predictions. Compared to $PM_{2.5}$ models, $NO_2$ predictions typically show more within-city variation. Models without UK show more within-city spatial variation for $PM_{2.5}$ than models with UK, however, this may be an artifact of local land use data serving as a proxy for explaining regional concentration variations. A similar phenomenon has been shown for $PM_{2.5}$ models with and without satellite data. [49] For $PM_{2.5}$ models, land-use variables tend to under-predict in high concentration areas like Beijing, and over-predict in relatively low concentration areas like Guangzhou. Satellite data made less of a contribution to our $NO_2$ models than to our $PM_{2.5}$ models. At a city scale, $PM_{2.5}$ models had a relatively low $R^2$ but a reasonable RMSE in most cities, which might be due to undetectable within-urban variability, or a lack of within-urban variability altogether. $NO_2$ models had better $R^2$ than $PM_{2.5}$ in most cities. Our city scale performance is comparable to some city scale models for Beijing ($R^2$ 0.78 vs. 0.58 for $PM_{2.5}$)[17] and Shanghai ($R^2$ 0.70 vs. 0.61 for $NO_2$)[19], but worse in some other locations (0.25 vs. 0.73 for $PM_{2.5}$ in Tianjin)[56]. Regional or city-scale models may better capture within-city variability than national models such as ours, especially for $PM_{2.5}$ for which pollution sources can be complex and vary by regions in China.

Since spatial resolutions of existing national-scale empirical models in China were 3km or larger, we also quantitatively computed modeled value variance within each $3\times3$ km$^2$ and $10\times10$ km$^2$ moving window based on $1\times1$ km$^2$ resolution predictions derived from our best performing models ($PM_{2.5}$-6 and $NO_2$-4, see Figure S12). $NO_2$ models consistently have much higher ratio than $PM_{2.5}$ models; that finding indicates that finer resolution predictions based on our models help to reveal with-city variations for $NO_2$ but less so for $PM_{2.5}$.

4.6 Limitations

A critical aspect of our approach is that it relies on regulatory monitors; such monitors often are located near specific land uses (e.g. public institutions, parks, traffic), but may not capture the full range land uses. Our approach uses a single model with fixed variable parameters to predict $PM_{2.5}$ and $NO_2$ for the whole country, however, the relationship between land uses and concentrations may vary by region. Incorporation of satellite data and universal kriging help to partially remedy this shortcoming. Finally, we did not incorporate kriging in variable selection in order to lighten the computational load, which may have resulted in underestimating the benefit of kriging.

**5. Conclusion**

We built national LUR models for ambient annual average $NO_2$ and $PM_{2.5}$ concentrations in China and generated publicly available $1\times1$ $km^2$ spatial resolution national prediction maps which could be used for national-scale long-term exposure analyses. Our models leverage information from ~900-1400 regulatory monitors, satellite-based measurements of $NO_2$ and $PM_{2.5}$, and 290 land use and meteorological variables. We find that parsimonious forward stepwise variable selection provides similar or better model performance than more computationally-intensive PLS variable reduction, an important finding for fine spatial resolution national predictions. We also find that categorized POI data from mapping services are a useful predictor in national scale LUR models and may provide information on local and regional sources that are not well captured from other nationally available data. Our models are capable of providing point predictions, such as at individual residential locations, which could be useful for other population-based environmental and environmental health studies in China, including in epidemiology, risk assessment, and environmental justice research. The general approach could usefully be applied to future years of data. Methodological findings here can inform future LUR research.

## Acknowledgments

This article was developed in part under Assistance Agreement no. RD83587301 awarded by the US Environmental Protection Agency (EPA). This article has not been formally reviewed by the US EPA. The views expressed in this document are solely those of the authors and do not necessarily reflect those of the agency. The US EPA does not endorse any products or commercial services mentioned in this publication. This research was also supported by Tsinghua Scholarship for Overseas Graduate Studies (No.2016143).